
\def\twa#1{\raise1.5ex\hbox{$\leftrightarrow$}\mkern-16.5mu #1}

\def\t{\tau}

\def\p{\prime}

\def\ria{\rightarrow}
\def\pp{\prime\prime}
\def\x{{\bf x}}

\def\s{\sigma}
\def\xpp{{\bf x}^{\prime\prime}}
\def\tpp{t^{\prime\prime}}
\def\Xpp{{x^{\mu}}^{\prime\prime}}
\def\X'{{x^{\mu}}^{\prime}}

\def\G{{\cal G}}
\def\R{{I\kern-0.3em R}}
\def\ula#1{\raise2ex\hbox{$\leftarrow$}\mkern-16.5mu #1}
\def\ura#1{\raise2ex\hbox{$\rightarrow$}\mkern-16.5mu #1}
\def\ulra#1{\raise1.5ex\hbox{$\leftrightarrow$}\mkern-16.5mu #1}
\def\X{x}

\def\Xpp{x''}
\def\pp{\prime\prime}
\def\p{\prime}

\def\x{{\bf x}}
\def\G{{\cal G}}
\def\C{{\cal C}}

\font\re=cmr8
\font\rn=cmr9

\font\sln=cmsl9
\def\narr{\advance\leftskip by 1.5 pc \advance\rightskip by 1.5 pc}

\magnification=\magstep1

\hfill CTP \# 2235
\par
\hfill gr-qc/9308020
\vskip 0.3 true in

\centerline{\bf QUANTUM MECHANICAL
COMPOSITION LAWS}
\centerline{{\bf IN REPARAMETERIZATION INVARIANT SYSTEMS}\footnote{*}{To
appear in the proceedings of {\sl Journ\'ees Relativistes 93}, World
Scientific, 1993.}}
\vskip 0.4 true in
\centerline{J. J. HALLIWELL\footnote{$^\dagger$}{e-mail:
j\_halliwell@vax1.physics.imperial.ac.uk}}
\vskip 0.02 true in
\centerline{\re Blackett Laboratory,
Imperial College of Science, Technology and Medicine,}
\centerline{\re Prince Consort Road, London SW7 2BZ,
UK.}
\vskip 0.2 true in
\centerline{M. E. ORTIZ\footnote{$^\ddagger$}{e-mail: ortiz@mitlns.mit.edu}}
\vskip 0.02 true in
\centerline{\re Center for Theoretical Physics, 6-412A,
Massachusetts Institute of Technology,}
\centerline{\re 77 Massachusetts Avenue,
Cambridge, MA 02139,
USA.}
\vskip 0.35 true in
\midinsert
\centerline{ABSTRACT}
\vskip 2pt
\re\baselineskip=10pt
\narr
This paper gives a brief description of the derivation of a composition law
for the propagator of a relativistic particle, in a sum over histories
quantization. The extension of this derivation to the problem of finding a
composition law for quantum cosmology is also discussed.
\endinsert
\vskip 0.15 true in
\baselineskip=13pt
It has been
argued that the sum over histories should be regarded as a more generally
applicable method of quantization than the canonical framework,
in the sense that the
former method may be applied to a broader variety of theories.$^1$
A case in point is general relativity, which admits, at least formally, a
sum over histories quantization, but may not admit a canonical
one. The existence of a composition
law is closely related to the derivation of
a canonical formalism from the sum over histories. This observation
provides the motivation for our discussion of the derivation of a
composition law for the relativistic particle propagator.
Although we know that in this particular case, composition laws, of the
form,
$$
	\G(\Xpp|\X') = -\int_\Sigma d\sigma^\mu \ \G(\Xpp|x)
		\ \twa{\partial_\mu}
		\ \G(x|\X') \ .
	\eqno(1)
$$
and
indeed a canonical formulation, exist, it is nevertheless of interest to
examine how these can be {\it derived}. This will allow us to examine
how the derivation
might or might not generalize to sums over histories in other
reparameterization invariant theories, such as quantum cosmology.
A more detailed treatment of the discussion given in this paper may be
found in Ref. 2, where proofs are provided for all of the statements below.
We shall not include further citations of Ref. 2.

In non-relativistic quantum mechanics, the propagator $g(\x'',t''|\x',t')$
is represented by a sum over histories
$$
	g(\x'',t''|\x',t')=\sum_{p(\x',t '\to \x'',t'')}\exp[iS(p)]
	\eqno(2)
$$
in which the paths $p$ move forwards in time $t$. The composition law
$$
	g(\x'',t''|\x',t')=\int d^3\x\, g(\x'',t''|\x,t)\, g(\x,t|\x',t')
	\eqno(3)
$$
then follows from the fact that each path intersects an intermediate
surface of constant time once only, and a {\it partition} of paths
according to their crossing position may be defined. Thus using (2),
(3) may be written as
$$
\eqalignno{
	&\sum_{\x_t} \ \sum_{p(\x',t' \ria \x_t,t)}
		\ \sum_{p(\x_t,t \ria \xpp,\tpp)}
		\ \exp\left[ iS(\x',t' \ria \x_t,t)
		+ iS(\x_t,t \ria \xpp,\tpp) \right]
	\cr
	&= \sum_{\x_t}
	g(\x'',t''|\x_t,t)\, g(\x_t,t|\x',t')\ .
	&(4)
	\cr
}
$$

In relativistic quantum mechanics, by contrast, the propagators may be
represented by sums over histories in which the paths go forwards or
backwards in time. This arises because the kinetic part of the
action for a relativistic particle
$$
	S=-m \ \int^{\t^{\p\p}}_{\t^\p}d\t \ \left[{\partial x^\mu\over
		\partial\t}{\partial x^\nu\over
		\partial\t}\eta_{\mu\nu}\right]^{1/2},
	\eqno(5)
$$
involves derivatives with respect to the parameter time $\tau$ rather than
the spacetime time, $t$. The reparameterization invariance of the action is
manifested by the fact that the propagator is independent of the initial
and final values of $\tau$. In fact, it takes the simple form
$$
	\G(x''|x')=\int dT\, g(x'',T|x',0)
	\eqno(6)
$$
where $T$ is proportional to $\tau''-\tau'$, and $g(x'',T|x',0)$ is the
quantum mechanical propagator with Hamiltonian $H=p^2-m^2$, and with $x''$
and $x'$ co-ordinates on $R^4$.
The integration
range of $T$ determines the boundary conditions of the propagator $\G$. For
example, integration from 0 to $\infty$ yields the $i$ times the
Feynman propagator $G_F$. This
expression for the propagator is known as the proper time representation,
and similar expressions exist for propagators in other
reparameterization invariant systems.

Since the paths summed over in $g(x'',T|x',0)$ are free to move forwards or
backwards in any co-ordinate time $x^0$, a general path intersects a
surface of constant time arbitrarily many times. As a consequence, a
relativistic version of the composition law (3) is not readily recovered
from this representation of the relativistic propagator. Below we
shall describe how a composition law may be derived, despite this apparent
difficulty.

Making use of the proper time representation, (6), where the path integral
has the form of a non-relativistic sum over histories, integrated over
time, the question of a composition law may be related to the question of
factoring the propagators of non-relativistic quantum mechanics across an
arbitrary surface in configuration space (in this case $R^4$). This may be
achieved using a known result called the Path Decomposition Expansion
(PDX).$^3$

The PDX is based on a partition of paths according to their {\it
first} crossing
position and time of a surface in configuration space, and may be stated as
follows. Let $g(\x'',T|\x',0)$ be a quantum
mechanical propagator from a point $\x'$ in a region $\C_1$ of
configuration space, to a point $\x''$ in a region $\C_2$ of configuration
space, and let $\Sigma$ be the boundary between $\C_1$ and $\C_2$. We
assume that the quantum mechanical system is derived from a Lagrangian
$L={1\over 2}M\dot{\x}^2 - V(\x).$
Then the PDX states that the propagator may be decomposed into the
composition of a restricted propagator in $\C_1$ from $(\x',0)$ to
$(\x_\sigma,t)$, composed with a standard unrestricted propagator in $\C_2$
from $(\x_\sigma,t)$ to $(\x'',T)$, with summations over both $\x_\sigma$ and
$t$, as
$$
	g(\x^{\pp}, T| \x', 0 ) = \int_0^T dt \int_{\Sigma} d\s
		\left.\ g(\x^{\pp}, T| \x_{\s} , t )
		\ {i \over 2M} \ {\bf n} \cdot {\bf \nabla}
		g^{(r)} (\x,t| \x', 0 ) \right\vert_{\x=\x_{\s}}\ .
\eqno(7)
$$
Here, $d\sigma$ is an integration over the surface $\Sigma$, and ${\bf n}$
is the unit normal to $\Sigma$ pointing into $\C_2$. The quantity
$g^{(r)}$ is the restricted propagator in $\C_1$. It can be defined as a
sum over paths lying only within $\C_1$, or as the propagator satisfying
the same differential equation as $g$, but with the condition that it
vanish on $\Sigma$.

It is also possible to partition the paths according to their {\it last}
crossing position and time, and this leads to a slightly different
composition law,
$$
	g(\x^{\pp}, T| \x', 0 ) = - \ \int_0^T dt \int_{\Sigma} d\s
		\ {i \over 2M} \ {\bf n} \cdot
		{\bf \nabla} \ g^{(r)} (\x^{\pp}, T| \x , t )
		\Big\vert_{\x=\x_{\s}}\ g (\x_{\s} ,t| \x', 0 )
\eqno(8)
$$
It is important to
emphasize that both (7) and (8) can be derived using only sum over paths
representations of $g$ and $g^{(r)}$.

Before we can make use of the PDX, it is necessary to eliminate the
restricted propagator $g^{(r)}$ from the expressions (7) and (8), since
we do not know how to relate it to the relativistic propagators $\G$.
This may be achieved by making use of a simple method of images type
argument$^4$ (which again may be derived from the sum over paths), which
states that provided that the potential $V(\x)$ satisfies obvious
symmetry properties (reflection symmetry about $\Sigma$),
$$
	{\bf n} \left.\left.\cdot {\bf \nabla} g^{(r)} (\x,t| \x', 0 )
		\right\vert_{\x=\x_{\s}}
		= 2 \ {\bf n} \cdot {\bf \nabla} g (\x,t| \x', 0 )
		\right\vert_{\x=\x_{\s}}\ .
\eqno(9)
$$
As a consequence, (7) becomes
$$
	g(\x^{\pp}, T| \x', 0 ) = \int_0^T dt \int_{\Sigma} d\s
		\left.\ g(\x^{\pp}, T| \x_{\s} , t )
		\ {i \over M} \ {\bf n} \cdot {\bf \nabla}
		g(\x,t| \x', 0 ) \right\vert_{\x=\x_{\s}}
\eqno(10)
$$
and (8) takes a similar form.

We are now ready to derive the composition law for the relativistic
propagator defined by (6). Let us concentrate on the Feynman Green
function where $T$ runs from $0$ to $\infty$. In this case, a
straightforward use of {\it both} (10) and its analogue derived from (8),
yields the familiar composition law,
$$
	G_F(x^{\pp}|x') = \ -\int_{\Sigma} d \s \ G_F(x^{\pp}|x)
		\ \ulra{\partial_n} \ G_F(x|x') \ ,
\eqno(11)
$$
where $\Sigma$ is a spacelike hypersurface in $R^4$.
The PDX can be used to derive composition laws for all
the relativistic propagators with proper time representations.

If we attempt to generalize the construction to a particle propagating
in a curved spacetime, we find that all the previous steps go through,
with the exception of the use of the method of images. This is only true
if the propagator is symmetric about each member of a family of factoring
surfaces $\Sigma$. A sufficient condition is that the metric have a
timelike Killing vector $\partial_t$, and that the chosen surfaces be
surfaces of constant $t$, {\it i.e.} the metric is static (we
anticipate that this will extend to stationary spacetimes). On an
arbitrary spacetime
a composition law cannot be derived. This observation should be
compared with the result$^5$
that a canonical formulation for a
quantized relativistic particle exists only if the ambient spacetime has
a timelike Killing vector.

Turning to quantum cosmology, where one may construct a closely
analogous propagator between three metrics, one can ask whether such an
object might satisfy a composition law. An important result$^5$
is that there is no Killing vector on superspace. We therefore
conclude that it is not possible to derive a composition law, and hence
a canonical formulation of quantum cosmology, using the techniques
outlined above. This result supports the notion that  the sum over
histories may be more general than the canonical approach to
quantization. However, it is important to emphasize that without a
canonical framework, there remains the important problem of
how the sum over histories may be used to
construct probabilities, {\it i.e.} the question of {\it interpretation}.
\vskip 0.2 true in

\noindent{\bf References}
\vskip 0.1 true in
{\rn
\item{1.} See for
example, J. B. Hartle, {\sln Phys. Rev.} {\bf D38} (1988) 2985;
{\sln Phys. Rev.} {\bf
D44} (1991) 3173; in {\sln Quantum Cosmology and Baby Universes},
Proceedings of the Jerusalem Winter School on Theoretical Physics, eds. S.
Coleman, J. Hartle, T. Piran and S. Weinberg (World Scientific, Singapore,
1991); in Proceedings of the {\sln International Symposium on Quantum Physics
and the Universe}, Waseda University, Tokyo, Japan (1992).
\item{2.} J. J. Halliwell and
	M. E. Ortiz, {\sln Phys. Rev.} {\bf D48} (1993) 748.
\item{3.} A. Auerbach and S. Kivelson, {\sln Nucl. Phys.} {\bf B257} (1985)
799.
\item{4.} L. Schulman and R. W. Ziolkowski, in {\sln Path Integrals form meV
to MeV}, eds V. Sa-yakanit, W. Sritrakool, J. Berananda, M. C. Gutzwiller,
A. Inomata, S. Lundqvist, J. R. Klauder and L. S. Schulman (World
Scientific, Singapore, 1989).
\item{5.} K. Kucha{\v r}, in {\sln Proceedings of the 4th Canadian
Conference on General Relativity and Relativistic Astrophysics}, eds G.
Kunstatter, D. Vincent and J. Williams (World Scientific, Singapore, 1992).
}
\end